\documentclass[twoside,twocolumn,9pt]{article}
\usepackage{extsizes}
\usepackage[super,sort&compress,comma]{natbib} 
\usepackage[version=3]{mhchem}
\usepackage[left=1.5cm, right=1.5cm, top=1.785cm, bottom=2.0cm]{geometry}
\usepackage{balance}
\usepackage{sectsty}
\usepackage{graphicx} 
\usepackage{lastpage}
\usepackage[format=plain,justification=justified,singlelinecheck=false,font={stretch=1.125,small,sf},labelfont=bf,labelsep=space]{caption}
\usepackage{float}
\usepackage{fancyhdr}
\usepackage{fnpos}
\usepackage[english]{babel}
\addto{\captionsenglish}{%
  \renewcommand{\refname}{Notes and references}
}
\usepackage{array}
\usepackage{droidsans}
\usepackage{charter}
\usepackage[T1]{fontenc}
\usepackage[usenames,dvipsnames]{xcolor}
\usepackage{setspace}
\usepackage[compact]{titlesec}
\usepackage{hyperref}

\usepackage{epstopdf}
\usepackage{amsmath}
\usepackage{amssymb}
\usepackage{mathtools}
\usepackage{bm}
\usepackage{indentfirst}
\usepackage{xcolor}

\graphicspath{{Figures/}}

\definecolor{cream}{RGB}{222,217,201}

\begin{document}

\pagestyle{fancy}
\thispagestyle{plain}
\fancypagestyle{plain}{
\renewcommand{\headrulewidth}{0pt}
}

\makeFNbottom
\makeatletter
\renewcommand\LARGE{\@setfontsize\LARGE{15pt}{17}}
\renewcommand\Large{\@setfontsize\Large{12pt}{14}}
\renewcommand\large{\@setfontsize\large{10pt}{12}}
\renewcommand\footnotesize{\@setfontsize\footnotesize{7pt}{10}}
\makeatother

\renewcommand{\thefootnote}{\fnsymbol{footnote}}
\renewcommand\footnoterule{\vspace*{1pt}%
\color{cream}\hrule width 3.5in height 0.4pt \color{black}\vspace*{5pt}} 
\setcounter{secnumdepth}{5}

\makeatletter 
\renewcommand\@biblabel[1]{#1}            
\renewcommand\@makefntext[1]%
{\noindent\makebox[0pt][r]{\@thefnmark\,}#1}
\makeatother 
\renewcommand{\figurename}{\small{Fig.}~}
\sectionfont{\sffamily\Large}
\subsectionfont{\normalsize}
\subsubsectionfont{\bf}
\setstretch{1.125} 
\setlength{\skip\footins}{0.8cm}
\setlength{\footnotesep}{0.25cm}
\setlength{\jot}{10pt}
\titlespacing*{\section}{0pt}{4pt}{4pt}
\titlespacing*{\subsection}{0pt}{15pt}{1pt}

\fancyfoot{}
\fancyfoot[LO,RE]{\vspace{-7.1pt}\includegraphics[height=9pt]{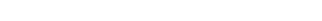}}
\fancyfoot[CO]{\vspace{-7.1pt}\hspace{13.2cm}\includegraphics{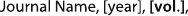}}
\fancyfoot[CE]{\vspace{-7.2pt}\hspace{-14.2cm}\includegraphics{head_foot/RF}}
\fancyfoot[RO]{\footnotesize{\sffamily{1--\pageref{LastPage} ~\textbar  \hspace{2pt}\thepage}}}
\fancyfoot[LE]{\footnotesize{\sffamily{\thepage~\textbar\hspace{3.45cm} 1--\pageref{LastPage}}}}
\fancyhead{}
\renewcommand{\headrulewidth}{0pt} 
\renewcommand{\footrulewidth}{0pt}
\setlength{\arrayrulewidth}{1pt}
\setlength{\columnsep}{6.5mm}
\setlength\bibsep{1pt}

\makeatletter 
\newlength{\figrulesep} 
\setlength{\figrulesep}{0.5\textfloatsep} 

\newcommand{\topfigrule}{\vspace*{-1pt}%
\noindent{\color{cream}\rule[-\figrulesep]{\columnwidth}{1.5pt}} }

\newcommand{\botfigrule}{\vspace*{-2pt}%
\noindent{\color{cream}\rule[\figrulesep]{\columnwidth}{1.5pt}} }

\newcommand{\dblfigrule}{\vspace*{-1pt}%
\noindent{\color{cream}\rule[-\figrulesep]{\textwidth}{1.5pt}} }

\makeatother

\twocolumn[
  \begin{@twocolumnfalse}
{\includegraphics[height=30pt]{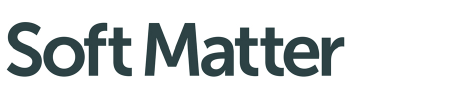}\hfill\raisebox{0pt}[0pt][0pt]{\includegraphics[height=55pt]{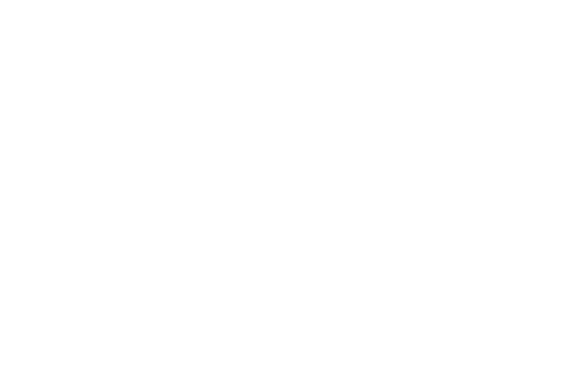}}\\[1ex]
\includegraphics[width=18.5cm]{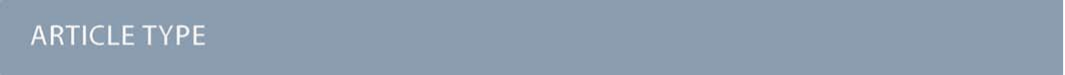}}\par
\vspace{1em}
\sffamily
\begin{tabular}{m{4.5cm} p{13.5cm} }

\includegraphics{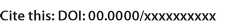} & \noindent\LARGE{\textbf{Two-dimensional diffusiophoretic colloidal banding: Optimizing the spatial and temporal design of solute sinks and sources$^\dag$}} \\
\vspace{0.3cm} & \vspace{0.3cm} \\

 & \noindent\large{Ritu R. Raj,\textit{$^{a}$} C. Wyatt Shields IV\textit{$^{a}$}\textit{$^{b}$} and Ankur Gupta$^{\ast}$\textit{$^{a}$}} \\

\includegraphics{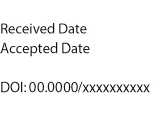} & \noindent\normalsize{Diffusiophoresis refers to the phenomenon where colloidal particles move in response to solute concentration gradients. Existing studies on diffusiophoresis, both experimental and theoretical, primarily focus on the movement of colloidal particles in response to one-dimensional solute gradients. In this work, we numerically investigate the impact of two-dimensional solute gradients on the distribution of colloidal particles, i.e., colloidal banding, induced via diffusiophoresis. The solute gradients are generated by spatially arranged sources and sinks that emit/absorb a time-dependent solute molar rate. First we study a dipole system, i.e., one source and one sink, and discover that interdipole diffusion and molar rate decay timescales dictate colloidal banding. At timescales shorter than the interdipole diffusion timescale, we observe a rapid enhancement in particle enrichment around the source due to repulsion from the sink. However, at timescales longer than the interdipole diffusion timescale, the source and sink screen each other, leading to a slower enhancement. If the solute molar rate decays at the timescale of interdipole diffusion, an optimal separation distance is obtained such that particle enrichment is maximized.  We find that the partition coefficient of solute at the interface between the source and bulk strongly impacts the optimal separation distance. Surprisingly, the diffusivity ratio of solute in the source and bulk has a much weaker impact on the optimal dipole separation distance. We also examine an octupole configuration, i.e., four sinks and four sources, arranged in a circle, and demonstrate that the geometric arrangement that maximizes enrichment depends on the radius of the circle. If the radius of the circle is small, it is preferred to have sources and sinks arranged in an alternating fashion. However, if the radius of the circle is large, a consecutive arrangement of sources and sinks is optimal. Our numerical framework introduces a novel method for spatially and temporally designing the banded structure of colloidal particles in two dimensions using diffusiophoresis and opens up new avenues in a field that has primarily focused on one-dimensional solute gradients.} \\

\end{tabular}

 \end{@twocolumnfalse} \vspace{0.6cm}

  ]

\renewcommand*\rmdefault{bch}\normalfont\upshape
\rmfamily
\section*{}
\vspace{-1cm}


\footnotetext{\textit{$^{a}$~Department of Chemical and Biological Engineering, University of Colorado, Boulder CO 80303, USA. E-mail: ankur.gupta@colorado.edu}}
\footnotetext{\textit{$^{b}$~Biomedical Engineering Program, University of Colorado Boulder, Boulder CO 80303, USA}}

\footnotetext{\dag~See DOI: 10.1039/cXsm00000x/}


\section{Introduction}
Diffusiophoresis is the phenomenon where colloidal particles move in response to solute concentration gradients.
The understanding of this key physical principle and its applications is enabling innovation in paint film deposition \cite{doi:10.1021/la00080a024}, laundry \cite{PhysRevApplied.9.034012}, membrane separation \cite{guha_diffusiophoresis_2015,florea_long-range_2014}, and hidden target searching \cite{tan_two-step_2021}. Solute concentration gradients in diffusiophoresis can be generated by a number of mechanisms \cite{velegol_origins_2016}: chemical reactions \cite{sharifi-mood_diffusiophoretic_2013}, mineral dissolution \cite{mcdermott_self-generated_2012}, and chemokine secretion \cite{kalinin_logarithmic_2009}, amongst others. The movement of colloidal particles due to concentration gradients can be divided into two broad categories: active and passive diffusiophoresis. In active diffusiophoresis \cite{brady_phoretic_2021,golestanian_propulsion_2005,singh_pystokes_2020,shaik_hydrodynamics_2021}, colloidal particles generate their own concentration gradients, while in passive diffusiophoresis \cite{migacz_diffusiophoresis_2022,singh_reversible_2020,abecassis_boosting_2008,ault_characterization_2019,hsu_diffusiophoresis_2017}, particles respond to an externally generated gradient.  

\par Recently, there have been numerous experimental and theoretical reports exploring the motion of active diffusiophoretic particles. These include the effects of finite Peclet numbers \cite{michelin_phoretic_2014,chang_diffusiophoresis_2019}, asymmetry in the form of Janus particles and bent rods  \cite{michelin_autophoretic_2015,venkateshwar_rao_self-propulsion_2019,ganguly_going_2022}, changes in the local fluid environment \cite{brady_phoretic_2021,shaik_hydrodynamics_2021,https://doi.org/10.1002/smll.200901976,doi:10.1021/la500182f}, and the use of active droplets instead of particles \cite{morozov_self-propulsion_2019,meredith_predatorprey_2020,izri_self-propulsion_2014}. Such systems have been proposed for uses in applications \cite{https://doi.org/10.1002/adma.201703660} such as environmental remediation \cite{gao_seawater-driven_2013}, drug delivery \cite{xuan_self-propelled_2014}, and cellular transport \cite{sanchez_controlled_2011}.

\par In contrast to active diffusiophoresis, there are several decades of literature on passive diffusiophoresis. One of the first series of studies to quantify the distribution of colloidal particles under diffusiophoresis was conducted by Staffeld et al. \cite{staffeld1989diffusion_nonelectro,staffeld1989diffusion_electro}. They showed, in electrolytic and non-electrolytic solutes, that the particle distribution exhibits a local maximum, resembling a band that moves with the diffusing solute front \cite{staffeld1989diffusion_nonelectro,staffeld1989diffusion_electro}. This laid the groundwork for studies of diffusiophoretic banding in other systems, including the well-studied dead-end pore geometry \cite{shi_droplet_2021,kar_enhanced_2015,singh_enhanced_2022,shim_diffusiophoresis_2022}. Experimental studies have been conducted on these dead-end pore systems to optimize nanoparticle transport in collagen hydrogels \cite{doi:10.1021/acs.nanolett.1c02251}, show the size dependence of particle transport into pores \cite{shin_size-dependent_2016}, determine design criteria for particle capture by a pore \cite{battat_particle_2019}, and develop a low cost zeta-potentiometer \cite{shin_low-cost_2017}. In addition to dead-end pore geometries, similar studies have been conducted in other microfluidic systems. Cross-channel pores have been used to study surface-solute interactions \cite{ault_characterization_2019} and the aggregration of colloidal particles near flow junctions \cite{shim_co_2020}. CO$_2$-induced concentration gradients across microfluidic channels have been used to predict exclusion zone formation in channel flows \cite{shim_co_2021}, remove bacteria from surfaces \cite{shim_co_2021_bact}, provide crossflow migration of colloids \cite{shimokusu_colloid_2020}, and enable membraneless water filtration \cite{shin_membraneless_2017}. In a similar way, salt gradients have been used to induce colloidal banding in microfluidic channels \cite{abecassis_boosting_2008,palacci_colloidal_2010}.

In addition to the breadth of experimental studies, analytical and numerical techniques have been used to study the phenomena observed in the aforementioned experimental systems. Anderson et al. showed that the diffusiophoretic velocity of a particle is dictated by surface interactions between the solute and particle \cite{anderson_colloid_nodate,prieve_motion_1984,anderson_motion_1982}. For ionic solutes, the diffusiophoretic velocity is given as $\bm{u}_{\rm{DP}} = M_{\rm{e}} \bm{\nabla} \ln c$, where $M_{\rm{e}}$ is the mobility of the particle and $c$ is the electrolyte concentration \cite{prieve_motion_1984}. For a particle moving in non-ionic solutes, the diffusiophoretic velocity is given as, $\bm{u}_{\rm{DP}} = M \bm{\nabla}c$, where $M$ is also a mobility parameter and $c$ is the solute concentration \cite{anderson_motion_1982}.  These mobility relationships can also be extended to include the effect of multiple ionic species \cite{alessio_diffusiophoresis_2021_2d,shi_diffusiophoretic_2016,chiang_multi-ion_2014}, arbitrary double layer thicknesses \cite{doi:10.1021/la991373k}, and ion sizes \cite{ohshima_ion_2022,stout_influence_2017}, amongst others.  Numerical studies have been conducted on the spreading of diffusiophoretic particles in response to applied solute gradients with hydrodynamic background flows \cite{chu_tuning_2022}, in one-dimensional transient gradients \cite{ault_diffusiophoresis_2017,chu_advective-diffusive_2020}, in concentrated electrolyte solutions \cite{gupta_diffusiophoresis_2020}, in solutes that exhibit Taylor dispersion due to a background/diffusioosmotic flow \cite{migacz_diffusiophoresis_2022,alessio_shim_gupta_stone_2022}, and in the presence of multiple electrolytes \cite{alessio_diffusiophoresis_2021_2d}.

\par Despite the expansive literature on passive diffusiophoresis, most studies focus on the effects of one-dimensional transient or steady solute concentration profiles on particle motion. The number of studies that expand particle motion to two or three dimensions are limited \cite{ault_diffusiophoresis_2018,ault_characterization_2019,battat_particle_2019,shim_co_2021,alessio_diffusiophoresis_2021_2d,migacz_diffusiophoresis_2022,alessio_shim_gupta_stone_2022,PhysRevLett.124.248004}, with most focusing on diffusiophoretic motion in two- and three-dimensional channel flows with one-dimensional driving solute gradients. 

\par Recently, Bannerjee et al. \cite{banerjee_soluto-inertial_2016} developed ``soluto-intertial" beacons that enable them to enact spatio-temporal control over solute gradients surrounding their beacons. This allows them to control and study diffusiophoretic particles moving in response to two- and three-dimensional gradients. They initially designed cylindrical hydrogel posts loaded with sodium dodecyl sulfate that attracted decane droplets and repelled polystyrene particles by releasing solute over a timescale of tens of minutes \cite{banerjee_soluto-inertial_2016}. By determining the appropriate diffusiophoretic velocity scale analytically in 3D and numerically in 2D, they were able to collapse the radial dependence of particle velocity \cite{banerjee_soluto-inertial_2016}. This proof-of-concept study showed that diffusiophoresis can be used as a mechanism to move colloidal particles deterministically over a length scale of hundreds of microns \cite{banerjee_soluto-inertial_2016}. The authors expanded this study to design temperature-triggered beacons, source and sink dipoles, dipoles with distinct solutes, and dipoles with reacting solutes \cite{banerjee_long-range_2019}. In follow-up studies, they developed design principles \cite{banerjee_design_2019}, which enabled them to manipulate colloidal distributions in suspension by a sedimenting beacon \cite{banerjee_drop-additives_2020} and deliver particles to hidden targets \cite{tan_two-step_2021}.

\par Inspired by the work from Banerjee et al. \cite{banerjee_long-range_2019} on source and sink dipoles, we envisioned that multiple solute sources and sinks can be spatially and temporally designed to optimize diffusiophoretic banding in two dimensions. To this end, we outline a numerical procedure for simulating diffusiophoretic colloidal transport in response to a non-electrolytic solute gradient generated by an arbitrary number of point sources and sinks. We determine an appropriate time-dependent molar rate by semi-analytically solving for the flux from a finite-sized solute source. Using our numerical scheme, we determine the timescales governing particle separation in a dipole and octupole source/sink system. For the dipole system, we show that there exists an optimum separation distance between the source and sink that maximizes particle enrichment in a specific region. This optimal distance is set by a balance between interdipole diffusion and molar rate decay timescales. We find that the optimal separation distance depends primarily on the partition coefficient, $K$, of the source/sink and is weakly dependent on the diffusivity ratio, $\hat{D}$. Lastly, we show how these principles change the optimal geometric arrangement of sources and sinks in an octupole configuration. Interestingly, we find that the optimal design of an octupole configuration depends on both the spatial arrangement of sources and sinks and the temporal decay of solute molar rate. These results underscore the rich dynamics observed by expanding diffusiophoretic driving forces to two dimensions. Our results also broaden the potential design space of colloidal banding using diffusiophoresis and provide a numerical framework to study the banding of diffusiophoretic particles in response to an arbitrary arrangement of solute sources and sinks.
\section{Problem setup}
\begin{figure}[t]
\centering
\includegraphics[width=\linewidth ]{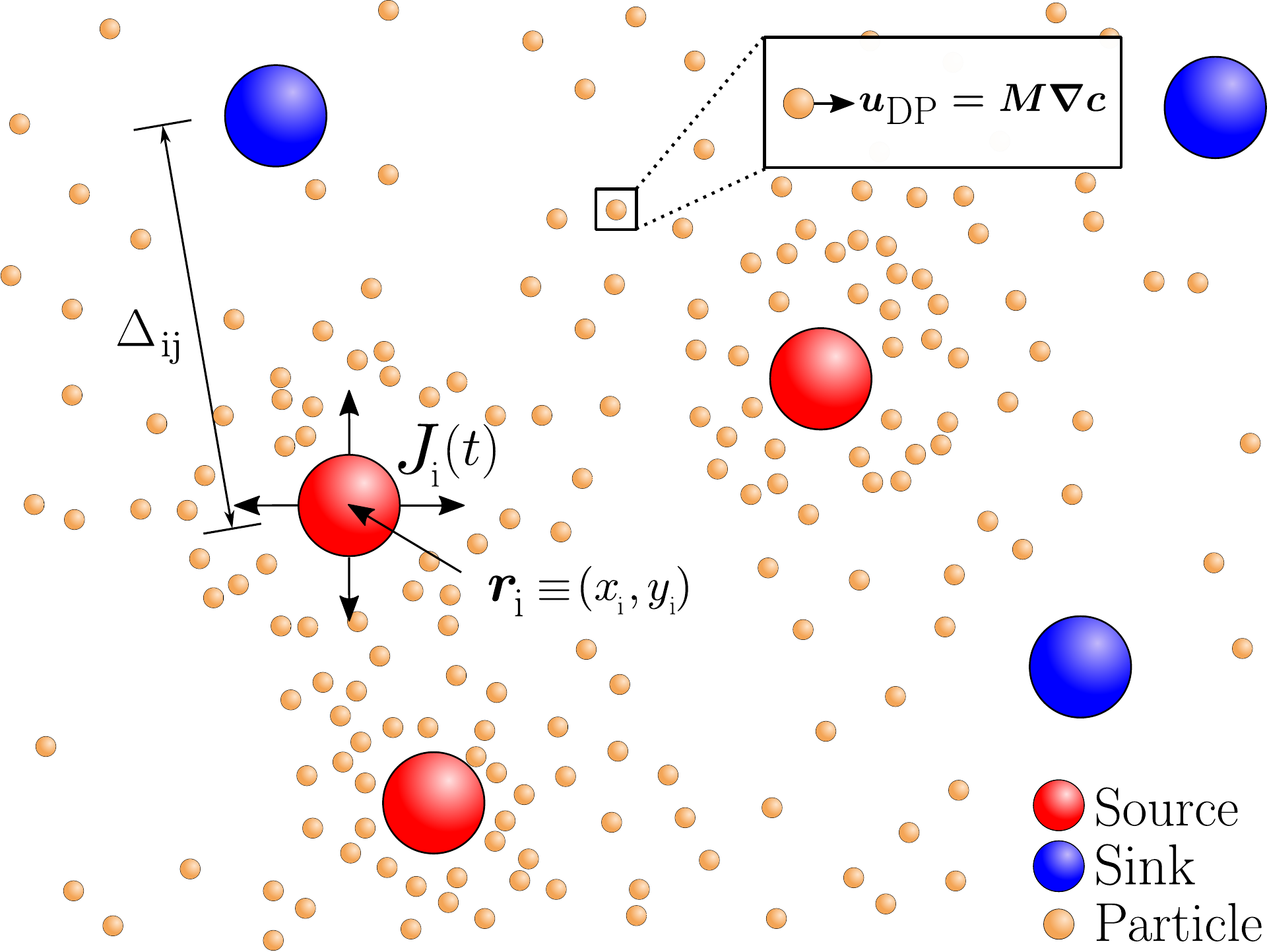}
\caption{\textbf{Schematic illustration of problem setup}. Solute sources and sinks are denoted by red and blue circles, respectively. The $\rm{i}^{\rm{th}}$ source/sink is located at a position $\bm{r}_{\rm{i}} \equiv \left(x_{\rm{i}},y_{\rm{i}}\right)$. The separation between the $\rm{i}^{\rm{th}}$ and $\rm{j}^{\rm{th}}$ source/sink is denoted as $\Delta_{\rm{ij}}$. The sources emit solute at a molar rate $J(t)$, whereas sinks absorb solute at a molar rate $-J(t)$. The emission and absorption of solute creates a concentration field, $c(\bm{r},t)$, which induces a diffusiophoretic velocity $\bm{u}_{\rm{DP}} = M\bm{\nabla}c$ on the particles, denoted by orange circles, where $M$ is the diffusiophoretic mobility.}
\label{fig1}
\end{figure}

To investigate the response of colloidal particles in two-dimensional solute gradients, described here as $\bm{\nabla} c$, we focus on the gradients generated by an arbitrary  number of solute sources and sinks. As shown in Fig. \ref{fig1}, we denote the locations of the sources and sinks by $\bm{r}_{\rm{i}}$, where the subscript $\rm{i}$ refers to the $\rm{i}^{\textrm{th}}$ source or sink. The distance between the $\rm{i}^{\textrm{th}}$ and $\rm{j}^{\textrm{th}}$ source or sink is denoted as $\Delta_{\rm{ij}}$. For simplicity, we consider that the sources emit solute at a molar rate $J(t)$ and that the sinks absorb solute at a molar rate of $-J(t)$. At time $t=0$, we have a uniform concentration of particles and solute in our system. At $t=0^{+}$, the sinks and sources begin emitting and absorbing the solute, creating a time-dependent and spatially varying concentration gradient. The solute gradient generated by sources and sinks induces a diffusiophoretic velocity on particles, $\bm{u}_{\rm{DP}} = M\bm{\nabla}c$. If $M > 0$, particles are attracted to the sources and repelled from the sinks. In contrast, if $M<0$, the particles are repelled from the sources and attracted to the sinks. At early times, the sources and sinks interact minimally, resulting in attraction/repulsion which transports particles towards the source and away from the sink (for $M > 0$). This creates local extrema of particle concentration, resulting in a banded distribution. As time progresses, the sources and sinks screen each other, much like electrostatic charges. At this timescale, the diffusiophoretic movement is diminished. In the following analysis, we seek to optimize particle enrichment by tuning the arrangement of sources and sinks, given a time-dependent molar rate, $J(t)$.
\par{} We acknowledge that in practical experimental setups, the emission and absorption rates are unlikely to be equal and opposite over time. However, while our numerical framework can handle arbitrary molar rates, we make this assumption to reduce the number of parameters in our system. In addition, we note that $\bm{u}_{\textrm{DP}}$ described above uses the non-electrolyte mobility relationship. The rationale to use this relationship is two-fold. First,  the non-electrolytic mobility expression does not possess the singularity found in the electrolytic mobility expression. We acknowledge that the singularity can be addressed by considering a concentration-dependent electrolytic mobility \cite{gupta_diffusiophoresis_2020,LEE2023130775}. For computational convenience, we refrain from incorporating a concentration dependent mobility relation. Second, if the concentration difference is relatively small, the two mobility relationships are equivalent; see Appendix \ref{Electrolytic and non-electrolytic mobilities for small concentration differences}. Therefore, we choose the non-electrolytic mobility relationship. We acknowledge that there might be quantitative differences if a different mobility relationship is employed, and comment on this difference in Appendix \ref{Electrolytic and non-electrolytic mobilities for small concentration differences}. Additionally, we acknowledge the limitation in using point sources and sinks, as spatial effects due to the presence of a finite-sized source/sink will yield differences. However, we  observe that the qualitative features remain the same as reported in prior experiments \cite{banerjee_long-range_2019}; see Appendix \ref{Qualitative comparison with experimental results}.
\subsection{Solute and particle transport equations}
 The species conservation equation for a solute concentration $c(\bm{r},t)$ is
\begin{equation}
    \frac{\partial c}{\partial t} = D_{\textrm{s}}\nabla ^2 c + \sum_{\textrm{i}}^{} J_{\textrm{i}} \left(t\right) \delta \left(\bm{r} - \bm{r}_{\textrm{i}}\right),
    \label{eq_solute_field}
\end{equation}
where $t$ is time, $D_{\textrm{s}}$ is the solute diffusivity, $\nabla$ is the gradient operator, $J_{\rm{i}}$ represents the strength of the $\rm{i}^{\rm{th}}$ source/sink, $\bm{r}$ is the position vector pointing from the origin, $\bm{r}_{\rm{i}}$ is the position of the $\rm{i}^{\rm{th}}$ source/sink and $\delta$ is the Dirac delta function. As is evident from eqn \eqref{eq_solute_field}, we treat solute sources and sinks as point sources. If the $\rm{i}^{\rm{th}}$ solute patch is a source, $J_{\rm{i}}=J(t)$, and if the $\rm{i}^{\rm{th}}$ solute patch is a sink, $J_{\rm{i}}=-J(t)$. As we show later, we account for the finite-size effect of the patch by deriving the emitted flux from an isolated source. We note that eqn \eqref{eq_solute_field} neglects any advection terms in solute transport, which is typical for studies on diffusiophoresis without background flows \cite{gupta_diffusiophoresis_2020,doi:10.1021/acs.langmuir.9b03333}. 
\par{} We calculate particle motion using two different approaches. First, we use Lagrangian particle tracking to determine the position of particles in time. The center of mass of the $\rm{i}^{\rm{th}}$ particle, $\bm{x}_{\rm{i}}$, can be determined by solving the following differential equation
\begin{equation}
\frac{d\bm{x}_{\rm{i}}}{dt} = \bm{u}_{\rm{DP}} =  M\bm{\nabla} c \big{|}_{\bm{x}_{\rm{i}}}.
\label{eq_part_tracking}
\end{equation}
We note that eqn \eqref{eq_part_tracking} neglects Brownian fluctuations. This is a typical assumption for diffusiophoretic particles as particle radii are typically $\mathcal{O}(10^{-6})$ m \cite{ault_diffusiophoresis_2018,alessio_shim_gupta_stone_2022}. 
\par{} Second, we calculate the concentration of colloidal particles, $n(\bm{r},t)$. The conservation equation for particle concentration is
\begin{equation}
\frac{\partial n}{\partial t} = D_{\rm{n}} \nabla ^2 n - \bm{\nabla}\cdot \left(n(M\bm{\nabla }c)\right),
\label{eq_part_continuum}
\end{equation}
where $D_{\rm{n}}$ is the diffusivity of the colloidal particles. The response of the particles to the generated solute field is included as an advective term. We retain $D_{\rm{n}}$ for numerical stability and assume $\frac{D_{\rm{n}}}{D_{\rm{s}}} \ll 1$. The retention of $D_{\rm{n}}$ helps  smooth the sharp gradients near the moving particle band. Eqns (\ref{eq_solute_field}) and (\ref{eq_part_tracking}) or eqns (\ref{eq_solute_field}) and (\ref{eq_part_continuum}) are solved simultaneously to determine $c(\bm{r},t)$, $\bm{x}_{\rm{i}}(\bm{r},t)$ and $n(\bm{r},t)$. 
\par {} Before numerically solving, we non-dimensionalize eqns (\ref{eq_solute_field})-(\ref{eq_part_continuum}) as
\begin{equation}
\frac{\partial \tilde{c}}{\partial \tau} = \tilde{\nabla} ^2 \tilde{c} + \sum_{\rm{i}}^{} \mathcal{J}_{\rm{i}} \left(\tau\right) \tilde{\delta} \left(\bm{\tilde{r}} - \bm{\tilde{r}}_{\rm{i}}\right),
\label{eq_solute_field_dim}
\end{equation}
\begin{equation}
\frac{d\bm{\tilde{x}}_{\rm{i}}}{d\tau} = \tilde{M}\bm{\tilde{\nabla} }\tilde{c}\big{|}_{\bm{\tilde{x}}_{\rm{i}}},
\label{eq_part_tracking_dim}
\end{equation}
\begin{equation}
\frac{\partial \tilde{n}}{\partial \tau} = \tilde{D} \tilde{\nabla} ^2 \tilde{n} - \bm{\tilde{\nabla}}\cdot \left(\tilde{n}(\tilde{M}\bm{\tilde{\nabla}} \tilde{c})\right),
\label{eq_part_continuum_dim}
\end{equation}
where $\mathcal{J}_{\rm{i}} = \frac{J_{\rm{i}}}{D_{\rm{s}}c_{\rm{ref}}}$, $\tilde{\delta} = \delta L^2$, $\tilde{n} = \frac{n}{n_{\rm{ref}}}$, $\tilde{c} = \frac{c}{c_{\rm{ref}}}$, $\tau = \frac{t}{L^2/D_{\rm{s}}}$, $\tilde{M} = \frac{Mc_{\rm{ref}}}{D_{\rm{s}}}$, $\tilde{D} = \frac{D_{\rm{n}}}{D_{\rm{s}}}$, $\tilde{\nabla} = L\nabla$, $\bm{\tilde{r}} = \frac{\bm{r}}{L}$, $\bm{\tilde{x}} = \frac{\bm{x}}{L}$, and $L$ is a reference length scale. We do not employ $a$ (i.e., the source/sink radius) or $\Delta$ as the reference length scale since $a$ only enters through our molar rate calculations and $\Delta$ is the variable that we seek to vary. We emphasize that $L$ is a reference length scale and does not influence our calculations. We solve these equations in a two-dimensional Cartesian domain with $\tilde{x},\tilde{y} \in [-10,10]$. We impose no-flux boundary conditions for both $\tilde{c}$ and $\tilde{n}$ on the domain boundaries. We set initial conditions $\tilde{n}(\bm{\tilde{r}},0) = \tilde{n}_0 = 1$ and $\tilde{c}(\bm{\tilde{r}},0) = \tilde{c}_0 = 0$. For simplicity, we take $\tilde{D} = 10^{-4}$ \cite{gupta_diffusiophoresis_2020, alessio_diffusiophoresis_2021_2d}. Additionally, we note that $\tilde{M} \leq 1$ for most colloids \cite{gupta_diffusiophoresis_2020} and use $\tilde{M} = 0.5$ for all simulations. To solve eqns (\ref{eq_solute_field_dim})-(\ref{eq_part_continuum_dim}), we need an input of $\mathcal{J}(\tau)$, which we discuss next. 
\par{} To elucidate the effects of molar rate decay, we use three different scenarios for $\mathcal{J}(\tau)$. First, constant molar rates, $\mathcal{J}(\tau) = \mathcal{J}_0\mathcal{H}(\tau)$, where $\mathcal{J}_0$ is the strength of the step molar rate and $\mathcal{H}$ is the heaviside function. In this scenario, there is no timescale associated with molar rate decay and the timescale for colloidal banding is dictated by the interaction between sources and sinks.  The second choice of $\mathcal{J}(\tau)$ is a boxcar function profile given by $\mathcal{J}(\tau) = \mathcal{J}_0\mathcal{H}(\tau)\mathcal{H}\left(\tau_0 - \tau\right)$, where $\tau_0$ introduces an additional timescale. 
\par{} Lastly, we derive $\mathcal{J}(\tau)$ by calculating the flux emitting from an isolated, finite-sized source of radius $a$. This allows us to incorporate experimentally relevant parameters, i.e., the partition coefficient of the solute into the source $K$, and the diffusivity ratio of solute between the source and the bulk $\hat{D}$. To evaluate $\mathcal{J}(\tau)$, we briefly restore dimensions. We assume the origin to be the center of the source. The inner region refers to the concentration field inside of the source, i.e., $r \le a$ and the outer region corresponds to the concentration field outside of the source, i.e., $r > a$. We assume that the concentration in the outer region is initially uniform  such that $c_{\rm{out}}=c_{\rm{ref}}$, and the source is saturated with solute such that the concentration in the inner region is $c_{\rm{in}} = K c_{\rm{ref}}$. At $t=0^{+}$, the concentration outside is switched to $c_{\textrm{out}}=0$, which leads the source to start emitting solute. The conservation equations for solute inside and outside the source are
\begin{equation}
\frac{\partial c_{\rm{in}}}{\partial t} = \frac{D_{\rm{in}}}{r} \frac{\partial}{\partial r}\left(r\frac{\partial c_{\rm{in}}}{\partial r}\right) \quad \quad r \le a,
\label{aux_in}
\end{equation}
\begin{equation}
\frac{\partial c_{\rm{out}}}{\partial t} = \frac{D_{\rm{s}}}{r} \frac{\partial}{\partial r}\left(r\frac{\partial c_{\rm{out}}}{\partial r}\right) \quad \quad r > a.
\label{aux_out}
\end{equation}
The initial and boundary conditions are
\begin{equation}
 \begin{dcases}
 &c_{\rm{in}}(r,t = 0) = Kc_{\rm{ref}} \\ 
 &c_{\rm{out}}(r,t = 0) = 0\\
 &\left. \frac{\partial c_{\rm{in}}}{\partial r}\right|_{r = 0} = 0\\
&c_{\rm{out}}(r \rightarrow \infty, t) = 0 \\
&c_{\rm{in}} (r = a,t) = Kc_{\rm{out}}(r = a,t)\\
&\-D_{\rm{in}} \left. \frac{\partial c_{\rm{in}}}{\partial r} \right|_{r = a} = \left. \-D_{\textrm{s}} \frac{\partial c_{\rm{out}}}{\partial r} \right|_{r = a}
\end{dcases} 
\end{equation}
We set the diffusivity of solute in the outer region to be the same as that of eqn (\ref{eq_solute_field}) and the diffusivity of the inner region to be $D_{\rm{in}}$. In order to determine the appropriate time dependence of flux from the source, we first non-dimensionalize the equations as follows:
\begin{equation}
    \frac{\partial \tilde{c}_{\rm{in}}}{\partial T} = \hat{D} \frac{1}{\tilde{r}} \frac{\partial}{\partial \tilde{r}} \left( \tilde{r}     \frac{\partial \tilde{c}_{\rm{in}}}{\partial \tilde{r}}       \right), \quad \tilde{r} < 1
    \label{aux_in_dim}
\end{equation}

\begin{equation}
    \frac{\partial \tilde{c}_{\rm{out}}}{\partial T} = \frac{1}{\tilde{r}} \frac{\partial}{\partial \tilde{r}} \left( \tilde{r}     \frac{\partial \tilde{c}_{\rm{out}}}{\partial \tilde{r}}       \right) , \quad \tilde{r} > 1
    \label{aux_out_dim}
\end{equation}
where  $\tilde{c}_{\rm{in}} = \frac{c_{\rm{in}}}{c_{\rm{ref}}}, \tilde{c}_{\rm{out}} = \frac{c_{\rm{out}}}{c_{\rm{ref}}},\hat{D} = \frac{D_{\rm{in}}}{D_{s}}$, $\tilde{r} = \frac{r}{a}$ and $T = \frac{\tau L^2}{a^2}$. We note that $T=\frac{\tau L^2}{a^2} = \frac{D_s t}{a^2}$ and is not influenced by $L$. By Laplace transforming the set of equations from $T$-space to $s$-space, we find a solution for the interfacial flux $\hat{F}(s)$; see Appendix \ref{Derivation of flux in the auxiliary problem}
\begin{equation}
    \hat{F}(s) = \frac{K\sqrt{\hat{D}}}{\sqrt{s}}\frac{K_{1,\rm{b}}(\sqrt{s})I_{1,\rm{b}}(\sqrt{\frac{s}{\hat{D}}})}{I_{0,\rm{b}}(\sqrt{\frac{s}{\hat{D}}})K_{1,\rm{b}}(\sqrt{s}) + K\sqrt{\hat{D}}I_{1,\rm{b}}(\sqrt{\frac{s}{\hat{D}}})K_{0,\rm{b}}(\sqrt{s})},
    \label{flux_s_space}
\end{equation}
where $I_{\rm{n,b}}$ and $K_{\rm{n,b}}$ are modified Bessel functions of the first and second kind, $\rm{n}^{\rm{th}}$ order. We numerically invert the flux from $s$-space to $T$-space, i.e. $F(T) =\mathcal{L}^{-1}\left(\hat{F}(s)\right)$, calculate the molar release rate, and appropriately scale the flux to get 
\begin{equation*}
\mathcal{J}(\tau) = 2\pi F \left(\frac{L^2}{a^2} \tau \right).
\end{equation*}
 $\mathcal{J}(\tau)$ is dependent on the partition coefficient $K$ and diffusivity ratio $\hat{D}$, which we discuss later.
 
\begin{figure*}[h]
\centering
\includegraphics[width=0.85\textwidth]{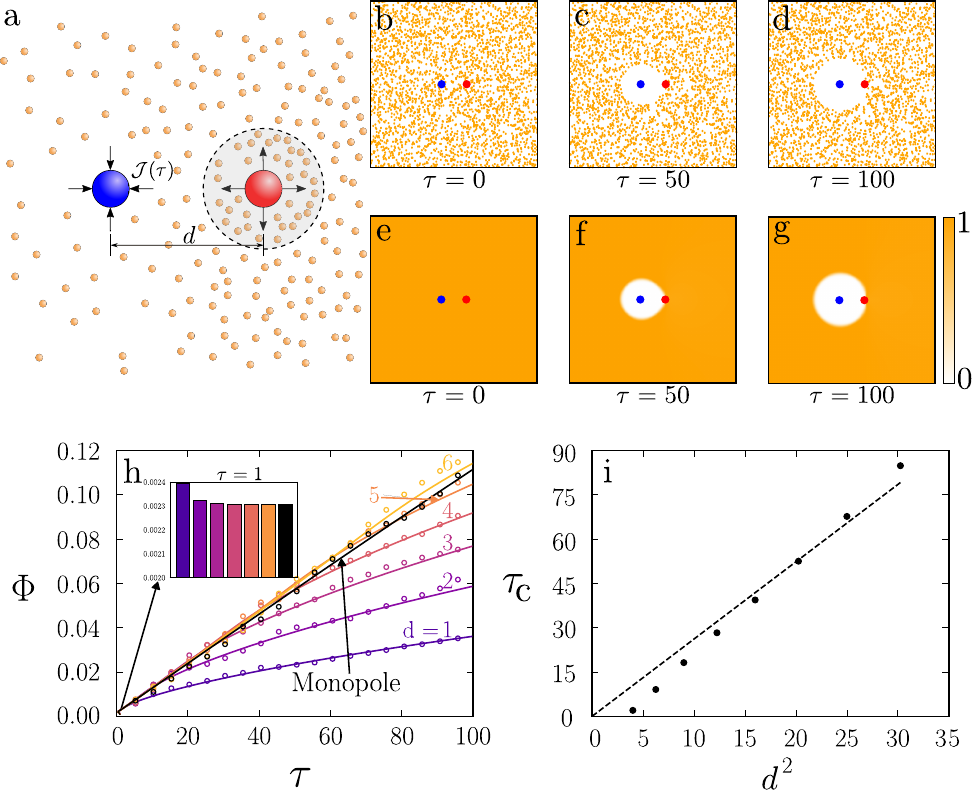}
\caption{\textbf{Dipole simulations for a constant molar rate.} (a) Schematic illustration of dipole setup where a source and a sink are separated by a distance $d$. The shaded region shows the $\Omega_1$ used in calculating $\Phi(\tau)$ via eqn (13) (b-d) $\tilde{\bm{x}}_i (\tau=0,50,100)$ for 3000 particles as calculated by solving eqns \eqref{eq_solute_field_dim} and \eqref{eq_part_tracking_dim} for $d = 3$ and $\tilde{M} = 0.5$. (e-g) $\tilde{n}(\bm{\tilde{r}}, \tau=0,50,100)$, as determined by solving eqns \eqref{eq_solute_field_dim} and \eqref{eq_part_continuum_dim} for $d=3$ and $\tilde{M} = 0.5$. The color bar ranges from 0 to 1. All concentration values larger than 1 are truncated to 1. (h) $\Phi(\tau)$ for a monopole and dipoles with $ d = 1 - 6$. Continuum results are represented with a solid line while particle tracking results are shown by open circles. Results for a source monopole are plotted in black. (h inset) $\Phi(\tau=1)$ for a monopole and dipoles with $ d = 1 - 6$ in the form of a bar chart. (i) $\tau_c$, i.e, the crossover time at which $\Phi(\tau)$ for the monopole overtakes a dipole with separation distance $d$, plotted versus $d^2$. The dotted line represents the line of best fit with zero intercept. $\mathcal{J}(\tau)= \mathcal{H}(\tau)$ for all panels.}
\label{fig2}
\end{figure*}

\subsection{Numerical schemes}
\textbf{Finite-volume method:}
To solve the coupled partial differential eqns (\ref{eq_solute_field_dim}) and (\ref{eq_part_continuum_dim}), we discretize both equations in space onto a square Cartesian grid with a grid size of 0.05 and write the resulting equations as coupled ordinary differential equations in time. We use a first-order upwinding scheme to resolve the convective term. We implement the point source/sink as a source term in the finite-volume cell, which contains the coordinates for the source/sink. For eqns (\ref{eq_solute_field_dim}) and (\ref{eq_part_tracking_dim}), we  discretize eqn (\ref{eq_solute_field_dim}) in space and solve the resulting equations with eqn (\ref{eq_part_tracking_dim}) as coupled ordinary differential equations in time. We interpolate the solute gradient at the position of the $\rm{i}^{\rm{th}}$ particle during each time step in order to determine the particle velocity. The coupled differential equations are then integrated using an eighth-order Runge-Kutta integration scheme (DOP853) as implemented in Scipy. To gain confidence in our simulations, we compare our results qualitatively to the experimental results of Banerjee et al. \cite{banerjee_long-range_2019} and obtain a good agreement; see Appendix \ref{Qualitative comparison with experimental results}.
\par{} \textbf{Optimization:} We define an objective function, which inputs the locations of sources and sinks for a given arrangement, solves eqns \ref{eq_solute_field_dim} and \ref{eq_part_continuum_dim} with a grid size of 0.1 and outputs a calculated fraction $\Phi(\tau)$. The fraction is defined as 
\begin{equation}
\Phi(\tau) = \frac{\displaystyle\int_{\Omega_1}\tilde{n}dV}{\displaystyle\int_{\Omega}\tilde{n}_0dV}.
\label{eq_fraction}
\end{equation}
\noindent $\Phi(\tau)$ represents the fractions of particles within a sub-region $\Omega_1$ of our domain $\Omega$. We employ the objective function into an optimization scheme to determine a source/sink arrangement that maximizes $\Phi(\tau)$. The optimization scheme uses a Nelder-Mead simplex algorithm implemented through the Scipy Optimization package.
\section{Results and discussion}
\begin{figure*}[h]
\centering
\includegraphics[width=0.8\textwidth]{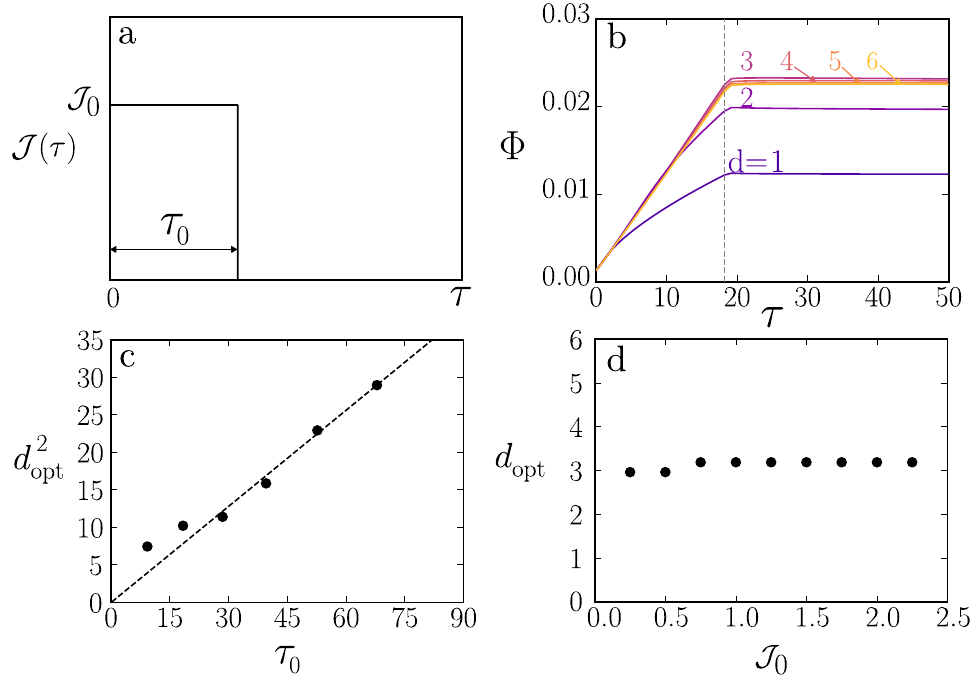}
\caption{\textbf{Effect of time-dependent molar rate on colloidal banding.}(a) Time-dependent source/sink molar rate profile described by the equation $\mathcal{J}(\tau) = \mathcal{J}_0\mathcal{H}(\tau)\mathcal{H}\left(\tau_0 - \tau\right)$, where $\mathcal H$ is the heaviside function. $\mathcal{J}_0$ is the strength of the molar rate and $\tau_0$ represents the time at which the source/sink molar rate vanishes. (b) $\Phi(\tau)$ for $d = 1-6$, $J_0=1$ and $\tau_0 = 18.2$. The vertical dotted line is placed at $\tau=\tau_0$. (c) $d_{\rm{opt}}^2$ versus $\tau_0$ for $\mathcal{J}_0=1$, where $d_{\rm{opt}}$ is the optimal separation distances, as estimated by our optimization scheme. (d) $d_{\rm{opt}}$ versus $\mathcal{J}_0$ for $\tau_0=18.2$.}
\label{fig3}
\end{figure*}
We begin our analysis with a dipole system, i.e., one source and one sink separated by a distance $d = \frac{\Delta}{L}$; see Fig. \ref{fig2}(a). The evolution of 3000 particle trajectories, as determined by eqns \eqref{eq_solute_field_dim} and \eqref{eq_part_tracking_dim}, for sources and sinks with constant strength $\mathcal{J}(\tau) = \mathcal{H}(\tau)$ and $\tilde{M}=0.5$ is provided in Fig. \ref{fig2}(b-d) (some representative contours for $\tilde{c}(\mathbf{r}, t)$ are provided in Appendix \ref{sec: c_field}). The evolution of particle concentration $\tilde{n}(\bm{\tilde{r}},\tau)$ for identical parameters as determined by eqns \eqref{eq_solute_field_dim} and \eqref{eq_part_continuum_dim} is displayed in Fig. \ref{fig2}(e-g). In both the particle and continuum simulations, since $\tilde{M}>0$, particles are repelled from the sink and are attracted to the source, forming a depletion zone around the sink and enrichment zone around the source. As time increases, particles enrich around the source and the depletion zone increases in size. To quantify enrichment, $\Phi(\tau)$ is calculated using eqn \eqref{eq_fraction}. We used a volume-averaged approach for quantifying enrichment as it is related to the enrichment phenomena observed experimentally \cite{banerjee_long-range_2019}. Fig. \ref{fig2}(h) shows that the fraction increases monotonically in time as particles enrich near the source. $\Phi(\tau)$ calculated with discrete and continuum simulations are in quantitative agreement.  Since the results of continuum simulations and particle tracking simulations are equivalent, for the remaining analysis, results from continuum simulations will be used. While the particle tracking simulations provide a descriptive picture of particle trajectories, they are computationally more expensive than continuum simulations since they require a large number of particles ($\sim$3000 in our analysis) to compute statistically significant volume averages.
\par{} Fig. \ref{fig2}(h inset) reveals that smaller $d$ values possess a higher $\Phi(\tau)$ for early times. In contrast, larger $d$ values display a higher $\Phi(\tau)$ at later times. We also compare these values with the enrichment from a single source, referred here as a monopole. At early times, the monopole provides the least enrichment, Fig \ref{fig2}(h inset). However, at long times, the monopole enrichment surpasses all dipoles. The time at which $\Phi(\tau)$ of the monopole overtakes $\Phi(\tau)$ of the dipoles is denoted as the crossover time, $\tau_c$. Fig. \ref{fig2}(i) shows a linear trend between $d^2$ and $\tau_c$. 
To explain the trends outlined above, we examine eqn \eqref{eq_part_continuum_dim} more carefully. First, we ignore diffusion as $\tilde{D} = 10^{-4}$. Next, we integrate eqn \eqref{eq_part_continuum_dim} over $\Omega_1$ (defined by the shaded region shown in Fig. \ref{fig2}a), and write
\begin{equation}
    \int_{\Omega_1}\frac{\partial \tilde{n}}{\partial \tau} dV = -\int_{\Omega_1} \bm{\tilde{\nabla}}\cdot \left(\tilde{n}(\tilde{M}\bm{\tilde{\nabla}} \tilde{c})\right)dV.
    \label{eq_derivation_dphidt}
\end{equation}
By employing eqn \eqref{eq_fraction} and divergence theorem, we obtain
\begin{equation}
    \frac{d \Phi}{d \tau} = -\frac{\tilde{M}}{N_0}\int_{S_1}\left(\tilde{n}\bm{\tilde{\nabla}} \tilde{c}\right)\cdot \hat{\bm{e}}_n dS,
    \label{eq_dphidt}
\end{equation}
where $N_0 = \int_{\Omega} \tilde{n}_0dV$, $S_1$ defines the outer perimeter of region $\Omega_1$, and $\hat{\bm{e}}_n$ is the unit normal vector pointing outwards from $S_1$. Essentially,  eqn \eqref{eq_dphidt} states that $\frac{d\Phi}{d\tau}$ is affected by the convective flux entering through $S_1$. The convective flux has two parameters,  i.e., $\tilde{\bm{\nabla}}\tilde{c}$ and $\tilde{n}$. 
\par{} At early times, dipoles have not had sufficient time to interact with each other. Therefore, we argue that to a first approximation, $\tilde{ \bm {\nabla}} \tilde{c}$ are similar for both a monopole and the source in dipoles. If so, to explain the trend in Fig. \ref{fig2}(h inset), eqn (\ref{eq_derivation_dphidt}) implies that at early times, $\tilde{n}$ is higher for smaller $d$ values. This appears surprising at first since the $\tilde{\nabla} \tilde{c}$ from sources and sinks do not interact at this timescale. However, the depletion of particles around the sink increases the concentration of particles at $S_1$, which consequently increases $\frac{d\Phi}{d\tau}$ (see Appendix \ref{dphidt for dipole simulations with constant flux}), leading to a larger $\Phi$.
\begin{figure*}[h]
\centering
\includegraphics[width=0.85\textwidth]{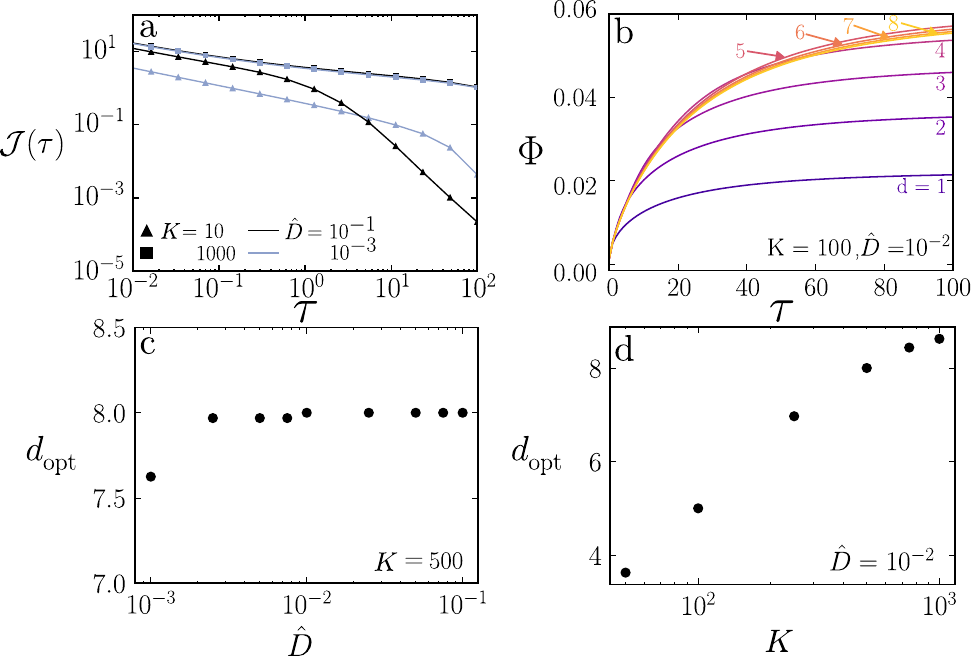}
\caption{\textbf{Optimal separation distance for experimentally realizable $\mathcal{J}(\tau)$.} (a) $\mathcal{J}(\tau)$, as calculated by inverting eqn \eqref{flux_s_space}, for a finite-sized source of radius $\frac{a}{L} = 0.4$. $K = 10, 1000$ and $\hat{D} = 10^{-1}, 10^{-3}$. (b) $\Phi(\tau)$ for $K = 100$ and $\hat{D} = 10^{-2}$, $d = 1-8$. (c) $d_{\rm{opt}}$ vs. $\hat{D}$ for $ K = 500$. (d) $d_{\rm{opt}}$ vs. $K$ for $\hat{D} = 10^{-2}$.}
\label{fig4}
\end{figure*}
\par{} We argue that dipoles start to interact with each other at $\tau \sim d^2$, or the interdipole diffusion time. For $\tau \gtrsim d^2$, the dipoles screen each other, causing a rapid decline in $\tilde{\nabla} \tilde{c}$. After the interdipole diffusion time, $\tilde{\nabla} \tilde{c}$ becomes localized between the source and sink and diminishes elsewhere. This results in a smaller $\frac{d\Phi}{d \tau}$; see eqn \eqref{eq_dphidt}. Since screening occurs later for larger $d$, the decay in $\frac{d \Phi}{d \tau}$ starts later and $\Phi(\tau)$ is higher; see Appendix \ref{dphidt for dipole simulations with constant flux}. Finally, for the monopole, screening never occurs, and concentration gradients do not diminish due to interactions with a sink. This is why the monopole overtakes dipoles around the interdipole diffusion time, which results in $\tau_c \sim d^2$; see Fig. \ref{fig2}(i). 
\par The aforementioned discussion highlights the time-dependent nature of enrichment. Therefore, we seek to study the effects of a time-dependent molar rate. To this end, we employ a molar rate profile given by $\mathcal{J}(\tau) = \mathcal{J}_0\mathcal{H}(\tau)\mathcal{H}\left(\tau_0 - \tau\right)$, where $\mathcal{H}$ is the Heaviside function;  see Fig. \ref{fig3}(a). This molar rate provides us with two parameters: the strength of the molar rate $\mathcal{J}_0$ and the time for the molar rate to decay to zero $\tau_0$. Fig \ref{fig3}(b) shows $\Phi(\tau)$ for $\mathcal{J}_0=1$, $\tau_0 = 18.2$ and $d=1-6$. The choice for $\tau_0$ corresponds to the crossover time observed in Fig. \ref{fig2} for $d=3$. For $\tau > \tau_0$ (represented by the dashed line in Fig. \ref{fig3}(b)), $\Phi(\tau)$ increases slightly before leveling. At $\tau=\tau_0$, we also observe that $\Phi(\tau)$ increases with separation distance until $d =3$ and then slightly decreases. Thus, there is an optimal separation distance. Using the described optimization scheme, we determined the optimal separation distance, $d_{\rm{opt}}$ as a function of $\tau_0$ and $\mathcal{J}_0$. In Fig. \ref{fig3}(c), we observe that a plot of $d_{\rm{opt}}^2$ versus $\tau_0$ results in a linear trend. Additionally, from Fig. \ref{fig3}(d) , we see that $d_{\rm{opt}}$ is weakly dependent on $\mathcal{J}_0$.
\par{} The $d_{\rm{opt}}$ is set by a balance between the interdipole diffusion and molar rate decay timescales. This is seen by the linear trend between $d_{\rm{opt}}^2$ and $\tau_0$ observed in Fig. \ref{fig3}(c). When $d < \sqrt{\tau_0}$, the source and sink screen each other before the molar rate is turned off, leading to small $\Phi(\tau)$. When $d \sim \sqrt{\tau_0}$, the enrichment around the source is boosted due to depletion around the sink, however, the source and sink do not screen each other as the molar rate vanishes at the inter-dipole diffusion time. Finally, when $d \gtrsim \sqrt{\tau_0}$, the enrichment around the source is less impacted by the depletion around the sink. In effect, $d \sim \sqrt{\tau_0}$ becomes the optimal distance. In summary, the timescale of molar rate decay can be used as a parameter to optimize particle enrichment. 
\begin{figure*}[h]
\centering
\includegraphics[width=0.85\textwidth]{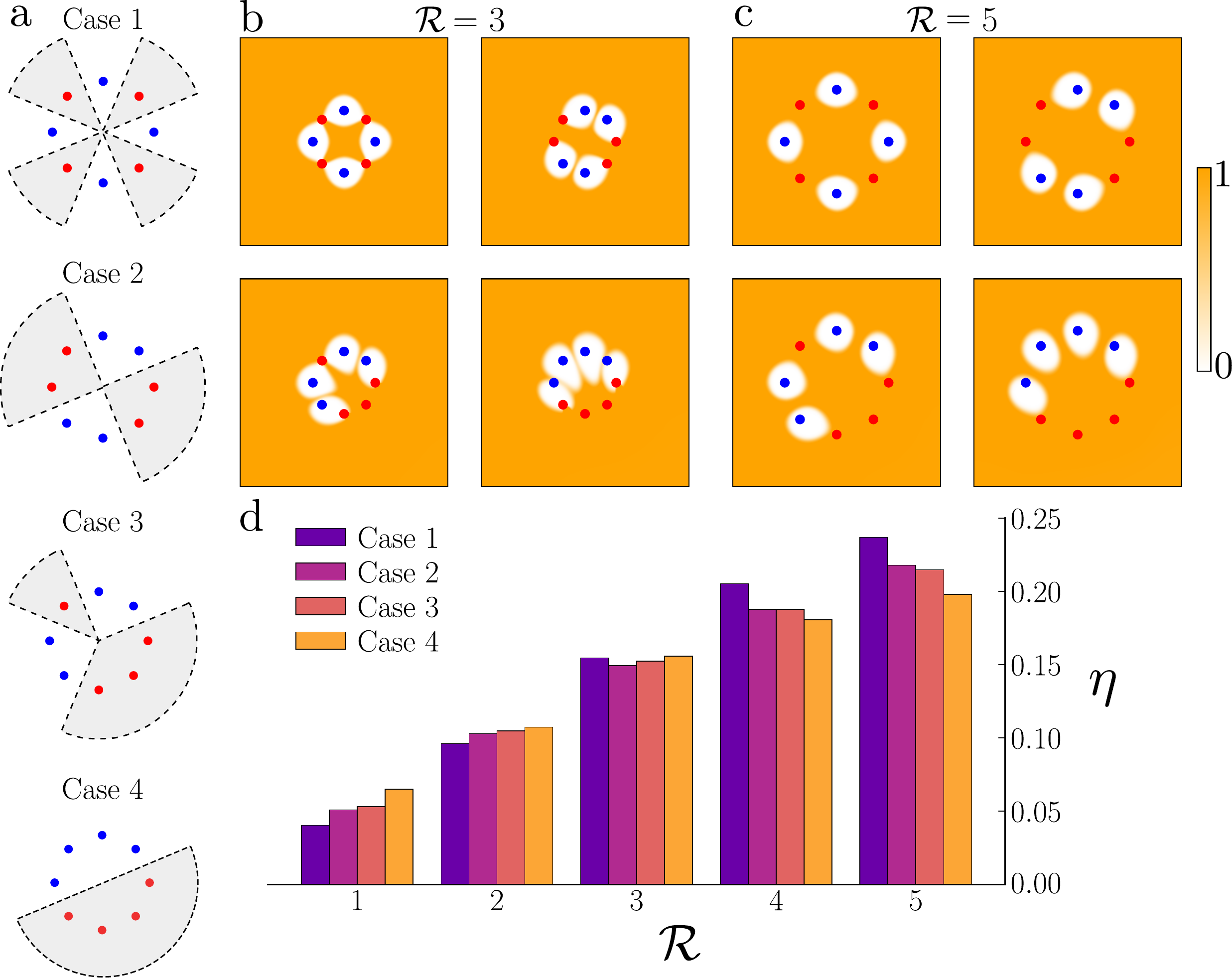}
\caption{\textbf{Geometric and spatial effects on banding for an octupole.} (a) Four arrangements studied in an octupole system with the shaded regions showing the $\Omega_1$ used in calculating $\Phi(\tau)$ via eqn \eqref{eq_fraction}. The sources and sinks are placed around a circle of radius $\mathcal{R}$. (b,c) Simulation snapshots at $\tau = 100$ with $\mathcal{J}(\tau) = \mathcal{H}(\tau)\mathcal{H}(18.2-\tau)$ for $\mathcal{R} = 3$ and $\mathcal{R} = 5$. (d) $\eta = \frac{\Phi(\tau = 100) - \Phi(\tau = 0)}{\Phi(\tau = 0)}$ for Cases 1-4 with $\mathcal{R}$ varying from $1-5$.}
\label{fig5}
\end{figure*}
\par $\mathcal{J}(\tau) = \mathcal{H}(\tau)$ and $\mathcal{J}(\tau) = \mathcal{J}_0\mathcal{H}(\tau)\mathcal{H}\left(\tau_0 - \tau\right)$ are not easy to realize experimentally. Instead, as shown by Banerjee et al. \cite{banerjee_soluto-inertial_2016, banerjee_design_2019,banerjee_long-range_2019}, solute fluxes arise due to solute partitioning between source and the bulk, described by a partition coefficient, denoted here as $K$. We also define the diffusivity ratio, $\hat{D}$, as the ratio of solute diffusivity in the source and in the bulk. As such, we incorporate the effects of these parameters by determining $\mathcal{J}(\tau) = f(K,\hat{D})$ using eqn \eqref{flux_s_space}. Fig. \ref{fig4}(a) shows $\mathcal{J}(\tau)$ for different values of $K$ and $\hat{D}$. As expected, the molar rate has a higher strength for a larger $K$ value, and the decay is slower for a smaller value of $\hat{D}$. 
\par{} We conduct dipole simulations by solving eqn \eqref{eq_part_continuum_dim} with $\mathcal{J}(\tau)$ determined by inverting eqn \eqref{flux_s_space}. We evaluate $\Phi(\tau)$ for different values for $K$ and $\hat{D}$. Fig. \ref{fig4}(b) shows $\Phi(\tau)$ with $\mathcal{J}(\tau) = f(K = 100, \hat{D} = 10^{-2})$ and for different $d$ values. Much like Fig. \ref{fig3}, we observe an optimal separation distance, $d_{\rm{opt}} \approx 5$. This demonstrates that $d_{\rm{opt}}$ is a generic feature of a time-dependent molar rate. We investigate the dependence of $d_{\rm{opt}}$ on $K$ and $\hat{D}$ using the optimization scheme described earlier. Fig. \ref{fig4}(c) shows the variation of $d_{\rm{opt}}$ with $\hat{D}$ for $K = 500$, where we observe that $d_{\rm{opt}}$ is weakly dependent on $\hat{D}$. However, Fig. \ref{fig4}(d) shows that $d_{\rm{opt}}$ is strongly dependent on $K$.  
\par {} The result of $d_{\textrm{opt}}$ showing a weak dependence on $\hat{D}$ is surprising, as one would expect $\hat{D}$ to impact the timescale of solute molar rate decay, which would ultimately impact the optimal separation distance. Therefore, we investigate this effect further. We note that there are two timescales for $\mathcal{J}(\tau) = f(K,\hat{D})$: a short timescale, during which solute transport occurs over a small boundary layer within the source, and a longer timescale where concentration gradients inside of the source are fully developed. An expansion of eqn \eqref{flux_s_space} around large $s$ (small $\tau$) shows that
\begin{equation}
\mathcal{J}(\tau) \sim \frac{K\sqrt{\hat{D}}}{(1 + K\sqrt{\hat{D}})\sqrt{\tau}}.
\label{eq_flux_shortime}
\end{equation}
Clearly, if the short timescale of molar rate decay balanced the interdipole diffusion timescale, then a dependence of $d_{\rm{opt}}$ on $\hat{D}$ would be observed. Interestingly, an expansion of eqn \eqref{flux_s_space} around small $s$ yields
\begin{equation}
    \hat{F}(s) \sim \frac{K}{2 + Ks\ln2 - \frac{K}{2}s\ln s}.
    \label{eq_longtime}
\end{equation}
Eqn \eqref{eq_longtime} is not analytically inverted, but we emphasize that it is only dependent on $K$. While an expansion for small $s$ cannot be directly related to large $\tau$, Fig. \ref{fig4}(d) shows that $d_{\rm{opt}}$ only depends on $K$. To this end, we argue that $d_{\rm{opt}}$ is determined by a balance between interdipole diffusion and long time scaling for $\mathcal{J}(\tau)$, which primarily depends on $K$. 

\par{} Given our understanding of timescales and their impact on optimal banding in dipole systems, we seek to expand our work to probe how the geometric arrangment of four sources and four sinks around a circle of radius $\mathcal{R}$, termed here as an octupole system, affects banding. Fig. \ref{fig5}(a) shows the four octupole arrangements we study. Case 1 refers to the arrangement where each source is nearest to two sinks and vice-versa, i.e., a relatively symmetric arrangement. Case 4 refers to the most asymmetric scenario where four sources are arranged consecutively, followed by four sinks. Case 2 and Case 3 are in between, with Case 2 being more symmetric than Case 3. The shaded areas outlined by dashed lines represent the integration region that $\Phi(\tau)$ is calculated over. Fig. \ref{fig5}(b,c) show simulation snapshots at $\tau = 100$ for $\mathcal{J}(\tau) = \mathcal{H}(\tau)\mathcal{H}(\tau_0 - \tau)$ with $\mathcal{R} = 3$ (panel b) and $\mathcal{R} = 5$ (panel c). $\tau_0=18.2$ is used for all simulations.
\par{} We quantify $\eta = \frac{\Phi(\tau = 100) - \Phi(\tau = 0)}{\Phi(\tau = 0)}$, i.e., the relative increase in $\Phi$. Fig. \ref{fig5}(d) shows $\eta$ for all four octupole arrangements, with $\mathcal{R}$ varying from $1$ to $5$. For $\mathcal{R} = 1$, Case 1 experiences the smallest increase in $\Phi(\tau)$, while Case 4 experiences the largest increase. As $\mathcal{R}$ increases from $1$ to $5$, this trend reverses and Case-1 experiences the largest increase in $\Phi(\tau)$ while Case 4 experiences the smallest increase. To understand this trend, we invoke our understanding from the dipole arrangement. The octupole has multiple interpole diffusion timescales. The smallest timescale is associated with $d_{\rm{ij}} = \frac{\Delta_{\rm{ij}}}{L} = 2 \mathcal{R} \sin \frac{\pi}{8}$ and the longest timescale is associated with $d_{\rm{ij}} = 2\mathcal{R}$. When $\mathcal{R} = 1$, the maximum $d_{\rm{ij}} \lesssim \sqrt{\tau_0}$. Therefore, all sources and sinks interact before the molar rate decays. In this scenario, the arrangement with the most geometric asymmetry, i.e., Case 4, has the largest $\eta$. Intuitively, in this case the source/sink screening is minimized, as the sources and sinks are collectively the furthest apart. When $\mathcal{R} = 5$, the smallest $d_{\rm{ij}} \gtrsim \sqrt{\tau_0}$, implying that none of the sources and sinks interact. Case 1 performs best in this regime, as sources are able to benefit from a local increase in $\tilde{n}(\bm{\tilde{r}},\tau)$ due to depletion from multiple nearby sinks. This effect is similar to the increase in performance for dipoles compared to a monopole observed earlier, see Fig. \ref{fig2}(h). Lastly, we  note that $\eta$, for all four cases, increases with $\mathcal{R}$ because $d_{\rm{ij}}$ also increases with $\mathcal{R}$. As $\mathcal{R}$ increases, the sinks and sources enrich particles for longer before interacting. We underscore that such complex banding patterns are unlikely to occur in one-dimensional diffusiophoretic systems as the motion of colloidal particles is  restricted to one direction.

\section{Conclusion}
In summary, we present a numerical framework for studying the banding of colloidal particles in response to two-dimensional concentration gradients. By studying the enrichment of particles in a dipole system, we find that both the interdipole diffusion and molar rate decay timescales impact the optimal banding of colloidal particles. Interestingly, a balance between these two characteristic timescales yields an optimal dipole separation distance, one which balances enrichment before the source and sink screen each other. By determining the flux from a finite-sized partitioning source, we include the effects of a partition coefficient $K$ and diffusivity ratio $\hat{D}$ into our molar rate profiles. We find that the optimal separation distance in this scenario depends primarily on $K$, with $\hat{D}$ only showing a weak effect. More importantly, we used the optimization of separation distance to elucidate that there are two timescales that impact the banding process. This discovery can be used to engineer complex systems with multiple sources and sinks. For instance, for an octupole arrangement of sources and sinks, we find that banding is also affected by geometric asymmetry. In fact, the optimal arrangement of sources and sinks is due to the interplay between multiple interpole diffusion timescales and the molar rate decay timescale. \par{} Looking forward, our results provide design principles for engineering microfluidic devices \cite{banerjee_design_2019,banerjee_long-range_2019,banerjee_soluto-inertial_2016,tan_two-step_2021} that utilize diffusiophoresis to move colloidal particles and create banded patterns. By utilizing partition coefficients and spatial arrangement, one can impart temporal and spatial control over the banded structure of colloidal particles. From a fundamental perspective, our results can also be expanded to include flow effects such as dispersion due to diffusiophoresis or diffusioosmosis \cite{shin_accumulation_2017,migacz_diffusiophoresis_2022,chu_tuning_2022,alessio_shim_gupta_stone_2022,singh_enhanced_2022,alessio_diffusiophoresis_2021_2d,doi:10.1021/acs.chemrev.1c00571}. Additionally, there is the potential to use such a system for applications that require precise control over colloid localization, such as biosensing \cite{squires_making_2008}, colloids separation \cite{shin_diffusiophoretic_2020}, and two-dimensional micropatterning \cite{choi_fabrication_2006}. Dipole and octupole systems, as envisioned, could be created using lithography similar to \cite{banerjee_long-range_2019}. Our work also invites future studies that move away from point sinks and sources, include higher-order effects and investigate asymmetric fluxes between sources and sinks. The results, as outlined in this article, motivate future experimental and theoretical studies to investigate two- and three-dimensional diffusiophoretic banding.

\appendix
\section{Electrolytic and non-electrolytic mobilities for small concentration differences}
\label{Electrolytic and non-electrolytic mobilities for small concentration differences}
The diffusiophoretic velocity for a particle moving in an electrolyte gradient can be written as
\begin{equation}
u_{\rm{DP}} = \frac{M_e}{c}\bm{\nabla}c.
\label{dp_elec_vel}
\end{equation}
If we consider a small concentration difference of the form $c(\bm{r},t) = c_0 + c_1(\bm{r},t)$, where $c_0$ is a constant concentration field and $c_1(\bm{r},t)$ is a small perturbation to that field such that $\frac{c_1}{c_0} \ll 1$, we can write eqn \ref{dp_elec_vel} as
\begin{equation}
u_{\rm{DP}} = \frac{M_e}{c_0 + c_1}\bm{\nabla}c_1 \approx \frac{M_e}{c_0}\bm{\nabla}c_1 = M_e'\bm{\nabla}c_1.
\end{equation}
For small concentration differences, the electrolytic and non-electrolytic diffusiophoretic velocities have the same form. We note that $M$ and $M_e'$ will have different values.
\par{} If the concentration difference is significant compared to the background concentration, the electrolytic and non-electrolytic expressions will yield a different response. Specifically, for an electrolytic mobility expression, the additional $\frac{1}{c}$ dependence will yield a higher $u_{\rm{DP}}$ around the sink. In contrast, $u_{\rm{DP}}$ will decrease around a source. We anticipate the qualitative features will remain the same. We invite interested readers to explore this effect quantiatively in future studies.
\section{Qualitative comparison with experimental results}
\label{Qualitative comparison with experimental results}

\begin{figure}[h]
\centering
\includegraphics[width=\linewidth]{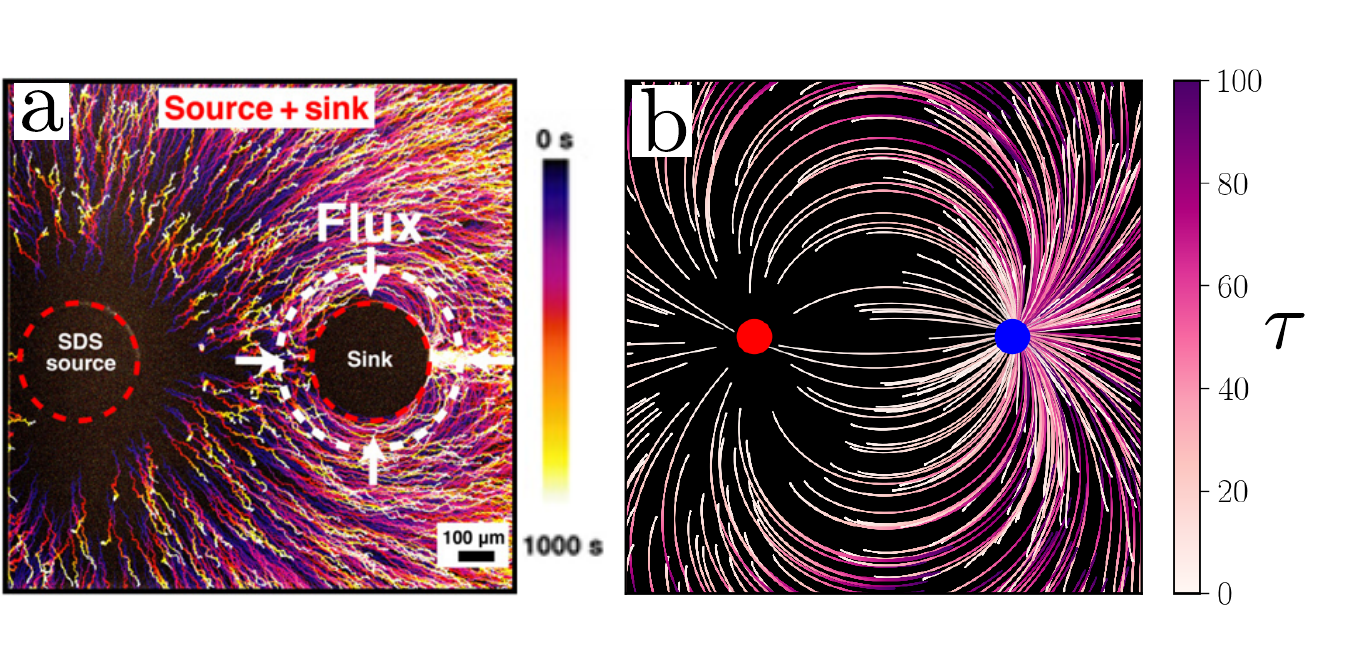}
\caption{\textbf{Comparison with experimental work by Banerjee et al. \cite{banerjee_long-range_2019}} (a) Example of particles moving in response to gradients generated from a source and sink, reproduced and adapted from \cite{banerjee_long-range_2019} with permission under a Creative Commons Attribution NonCommercial License 4.0 (CC BY-NC). (b) Particle streaklines showing time-coded trajectories for particles with $\tilde{M} = -0.5$. $d = 3$ and $\mathcal{J}(\tau) = \mathcal{H}(\tau)$. Simulation results are for $\tilde{x},\tilde{y} \in [-10,10]$, but are zoomed in to $\tilde{x},\tilde{y} \in [-3,3]$.}
\label{app_fig}
\end{figure}

\par{} We observe qualitative agreement with the work by Banerjee et al. \cite{banerjee_long-range_2019}. If $\tilde{M} = -0.5$, we see that particles move from the source towards the sink, Fig. \ref{app_fig}(b), similar to that observed in Fig. \ref{app_fig}(a). Additionally, as shown by the streaklines, we observe particles moving towards the side of the sink farthest from the source, similar to that observed in Fig. \ref{app_fig}(a). The observed qualitative agreement with experimental observations highlights the potential for our system to be used as a design tool in two-dimensional banding systems.

\section{Derivation of flux in the auxiliary problem}
\label{Derivation of flux in the auxiliary problem}
We Laplace transform  eqns (\ref{aux_in_dim}) and (\ref{aux_out_dim}) from $T$ to $s$-space as
\begin{equation}
    \mathcal{L} \left(  \frac{\partial \tilde{c}_{\rm{in}}}{\partial T} - \hat{D} \frac{1}{\tilde{r}} \frac{\partial}{\partial \tilde{r}} \left( \tilde{r}     \frac{\partial \tilde{c}_{\rm{in}}}{\partial \tilde{r}}       \right) \right) = s\hat{c}_{\rm{in}} - K = \hat{D} \frac{1}{\tilde{r}} \frac{\partial}{\partial \tilde{r}} \left( \tilde{r}     \frac{\partial \tilde{c}_{\rm{in}}}{\partial \tilde{r}} \right)
    \label{lap_in}
\end{equation}
\begin{equation}
    \mathcal{L} \left( \frac{\partial \tilde{c}_{\rm{out}}}{\partial T} - \frac{1}{\tilde{r}} \frac{\partial}{\partial \tilde{r}} \left( \tilde{r}     \frac{\partial \tilde{c}_{\rm{out}}}{\partial \tilde{r}}       \right)\right) = s\hat{c}_{\rm{out}} - 0 = \frac{1}{\tilde{r}} \frac{\partial}{\partial \tilde{r}} \left( \tilde{r}     \frac{\partial \tilde{c}_{\rm{out}}}{\partial \tilde{r}}       \right).
    \label{lap_out}
\end{equation}

\noindent We drop the tildes for convenience. We now have a set of two ordinary differential equations. We substitute $H = \hat{c}_{\rm{in}} - \frac{K}{s}$ in eqn (\ref{lap_in}) and obtain
\begin{equation}
    H = \frac{\hat{D}}{s} \frac{1}{r}\frac{\partial}{\partial r}\left(r\frac{\partial H}{\partial r}\right).
\end{equation}
Applying the product rule, we obtain the modified Bessel's equation
\begin{equation}
    r^2 \frac{\partial ^2 H}{\partial r^2} + r\frac{\partial H}{\partial r} - r^2 \frac{s}{\hat{D}}H = 0,
\end{equation}

\noindent which has a solution of the form
\begin{equation}
H = A(s)I_{0,\rm{b}}(\sqrt{\frac{s}{\hat{D}}}r) + B(s)K_{0,\rm{b}}(\sqrt{\frac{s}{\hat{D}}}r),
\end{equation}
where $I_{0,\rm{b}}$ and $K_{0,\rm{b}}$ are the zeroth-order modified Bessel functions of the first and second kind, respectively. Writing in terms of $\hat{c}_{\rm{in}}$, we get
\begin{equation}
    \hat{c}_{\rm{in}} = A(s)I_{0,\rm{b}}(\sqrt{\frac{s}{\hat{D}}}r) + B(s)K_{0,\rm{b}}(\sqrt{\frac{s}{\hat{D}}}r) + \frac{K}{s}.
\end{equation}
Applying the symmetry boundary condition at $r = 0$, we obtain that $B(s) = 0$ as $K_{0,\rm{b}} \rightarrow \infty$ when $r \rightarrow 0$. Thus, our solution for the inner problem in Laplace space reads
\begin{equation}
     \hat{c}_{\rm{in}} = A(s)I_{0,\rm{b}}(\sqrt{\frac{s}{\hat{D}}}r) + \frac{K}{s}.
\end{equation}

\noindent $A(s)$ will be determined when applying the partition and flux-matching boundary conditions. Returning to the outer problem, we write eqn (\ref{lap_out}) in terms of a modified Bessel's equation
\begin{equation}
    r^2 \frac{\partial^2\hat{c}_{\rm{out}}}{\partial r^2} + r\frac{\hat{c}_{\rm{out}}}{\partial r} - r^2s\hat{c}_{\rm{out}} = 0,
\end{equation}
which has a solution of the form
\begin{equation}
    \hat{c}_{\rm{out}} = M(s)I_{0,\rm{b}}(\sqrt{s}r) + G(s)K_{0,\rm{b}}(\sqrt{s}r).
\end{equation}
Applying the far field decay condition, $M(s)$ must be zero because $I_{0,\rm{b}} \rightarrow \infty$ as $r\rightarrow \infty$. Our solution to the outer problem is
\begin{equation}
\hat{c}_{\rm{out}} = G(s)K_{0,\rm{b}}(\sqrt{s}r).
\end{equation}
To determine our unknown coefficients, we apply the partition and flux matching boundary conditions. Starting with the partition boundary condition,
\begin{equation}
    A(s)I_{0,\rm{b}}(\sqrt{\frac{s}{\hat{D}}}) + \frac{K}{s} = KG(s)K_{0,\rm{b}}(\sqrt{s}).
\end{equation}
We solve for $G(s)$ and obtain
\begin{equation}
    G(s) = \frac{A(s)}{K}\frac{I_{0,\rm{b}}(\sqrt{\frac{s}{\hat{D}}})}{K_{0,\rm{b}}(\sqrt{s})} + \frac{1}{sK_{0,\rm{b}}(\sqrt{s})}.
    \label{partition}
\end{equation}
By applying the flux-matching condition, we write
\begin{equation}
    \hat{D}A(s)\sqrt{\frac{s}{\hat{D}}}I_{1,\rm{b}}(\sqrt{\frac{s}{\hat{D}}}) = -G(s)\sqrt{s}K_{1,\rm{b}}(\sqrt{s}),
    \label{flux_match}
\end{equation}
we solve for $A(s)$ by substituting eqn (\ref{partition}) into (\ref{flux_match}) to obtain 
\begin{equation}
    A(s) = \frac{-KK_{1,\rm{b}}(\sqrt{s})}{sI_{0,\rm{b}}(\sqrt{\frac{s}{\hat{D}}})K_{1,\rm{b}}(\sqrt{s}) + sK\sqrt{\hat{D}}I_{1,\rm{b}}(\sqrt{\frac{s}{\hat{D}}})K_{0,\rm{b}}(\sqrt{s})}.
\end{equation}
$G(s)$ is thus given by
\begin{equation}
    G(s) = \frac{K\sqrt{\hat{D}}I_{1,\rm{b}}(\sqrt{\frac{s}{\hat{D}}})}{sI_{0,\rm{b}}(\sqrt{\frac{s}{\hat{D}}})K_{1,\rm{b}}(\sqrt{s}) + sK\sqrt{\hat{D}}I_{1,\rm{b}}(\sqrt{\frac{s}{\hat{D}}})K_{0,\rm{b}}(\sqrt{s})}.
\end{equation}

\noindent We write our expression for $\hat{c}_{\rm{in}}$ and $\hat{c}_{\rm{out}}$ as

\begin{equation}
\hat{c}_{\rm{in}} = \frac{K}{s}\left(1 - \frac{K_{1,\rm{b}}(\sqrt{s})I_{0,\rm{b}}(\sqrt{\frac{s}{\hat{D}}}r)}{K_{1,\rm{b}}(\sqrt{s})I_{0,\rm{b}}(\sqrt{\frac{s}{\hat{D}}}) + K\sqrt{\hat{D}}K_{0,\rm{b}}(\sqrt{s})I_{1,\rm{b}}(\sqrt{\frac{s}{\hat{D}}})}\right),
\end{equation}

\begin{equation}
\hat{c}_{\rm{out}} =  \frac{K\sqrt{D}K_{0,\rm{b}}(\sqrt{s}r)I_{1,\rm{b}}(\sqrt{\frac{s}{\hat{D}}})}{sK_{1,\rm{b}}(\sqrt{s})I_{0,\rm{b}}(\sqrt{\frac{s}{\hat{D}}}) + sK\sqrt{\hat{D}}K_{0,\rm{b}}(\sqrt{s})I_{1,\rm{b}}(\sqrt{\frac{s}{\hat{D}}})}.
\end{equation}

Lastly, we find an analytical expression for the flux $\left( \left. \hat{F}(s) = -\frac{\partial \hat{c}_{\rm{out}}}{\partial r} \right |_{r = 1} \right)$ at the interface between the inner and outer region as

\begin{equation}
    \hat{F}(s) = \frac{K\sqrt{\hat{D}}}{\sqrt{s}}\frac{K_{1,\rm{b}}(\sqrt{s})I_{1,\rm{b}}(\sqrt{\frac{s}{\hat{D}}})}{I_{0,\rm{b}}(\sqrt{\frac{s}{\hat{D}}})K_{1,\rm{b}}(\sqrt{s}) + K\sqrt{D}I_{1,\rm{b}}(\sqrt{\frac{s}{\hat{D}}})K_{0,\rm{b}}(\sqrt{s})}.
\end{equation}

\section{$\frac{d\Phi}{d\tau}$ for dipole simulations with a constant molar rate}
\label{dphidt for dipole simulations with constant flux}
\begin{figure}[h]
\centering
\includegraphics[width=0.95\linewidth]{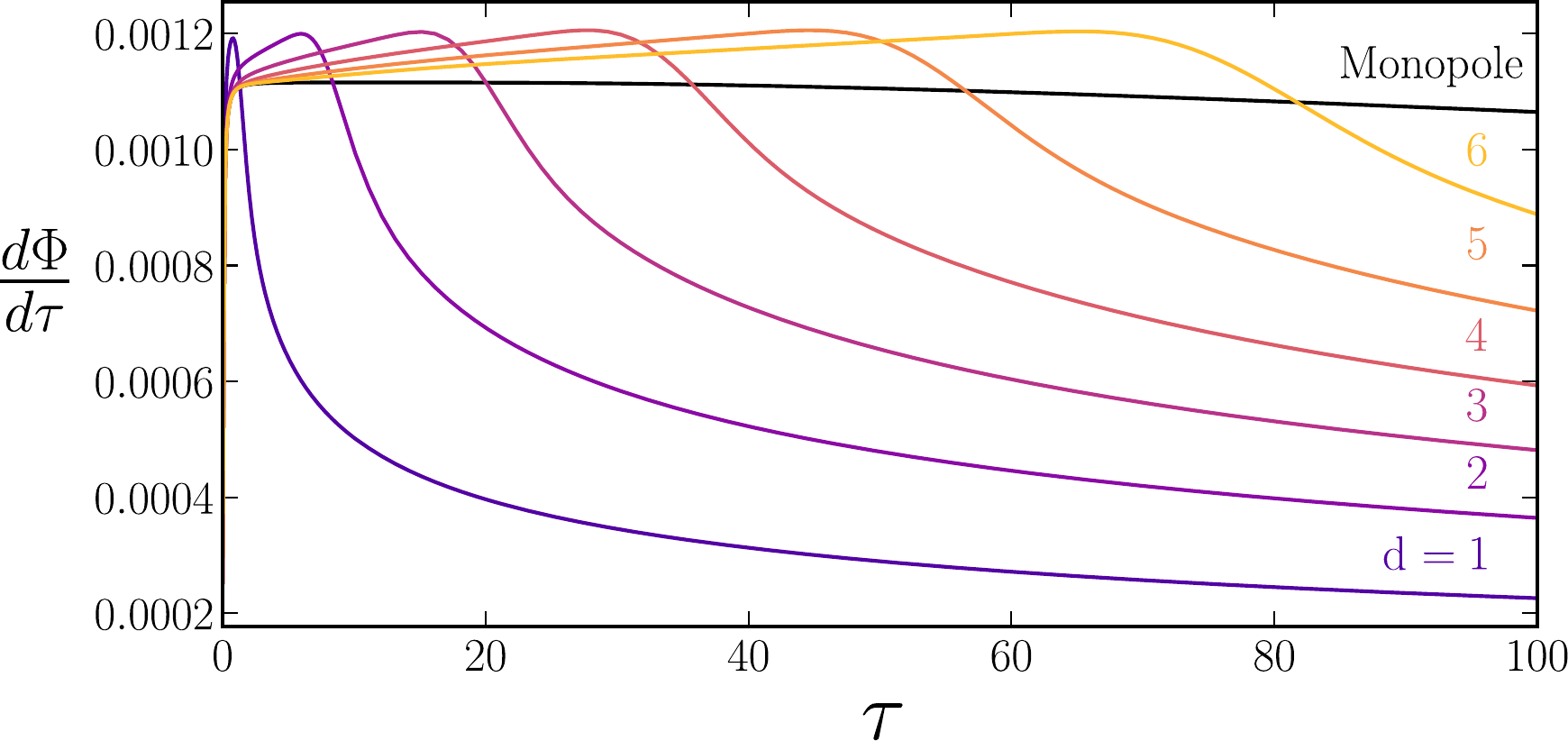}
\caption{\textbf{$\frac{d\Phi}{d\tau}$ for $\tilde{M} = 0.5$, $\mathcal{J}(\tau) = \mathcal{H}(\tau)$}. $\frac{d\Phi}{d\tau}$ for a monopole (black line) and dipoles with $d = 1-6$ for a constant molar rate $\mathcal{J}(\tau) = \mathcal{H}(\tau)$.}
\label{app_fig_dphidt}
\end{figure} 
We also calculate $\frac{d\Phi}{d\tau}$ for the dipole simulations with a constant molar rate; see Fig. \ref{app_fig_dphidt}. Initially the dipoles have larger $\frac{d\Phi}{d\tau}$, however, eventually $\frac{d\Phi}{d\tau}$ starts to decay. We argue that the initial increase in $\frac{d\Phi}{d\tau}$ is caused by enrichment at  $S_1$ due to depletion from the sink. We observe that $\frac{d\Phi}{d\tau}$ decays later for larger $d$. As the decay at longer times is caused by interactions between the sources and sinks, dipoles separated farther apart  screen each other later.
\section{Solute concentration field for a dipole with $d = 3$}
\label{sec: c_field}
\begin{figure}[h]
\centering
\includegraphics[width=\linewidth]{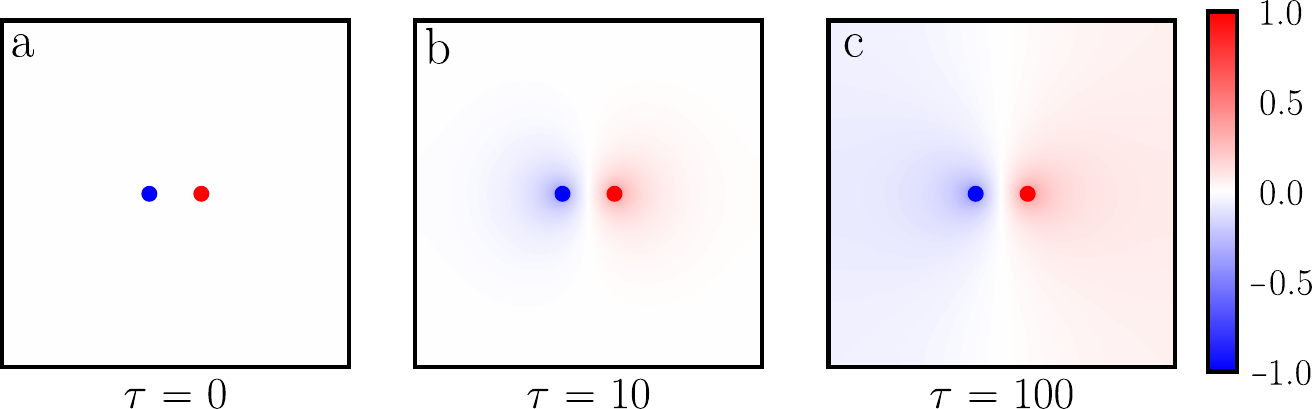}
\caption{\textbf{Concentration field generated by a point source and sink dipole}. (a-c) $\tilde{c}(\tilde{\bm{r}},\tau = 0,10,100)$ for a dipole with $d = 3$ and a molar rate $\mathcal{J}(\tau) = \mathcal{H}(\tau)$. The color bar ranges between -1 and 1 and represents the value of $\tilde{c}(\tilde{\bm{r}},\tau)$.  The point source and sink are visualized as a red and blue circle and are not representative of solute concentration at the location of the source and sink.}
\label{app_fig_conc_d_3}
\end{figure}
Fig. \ref{app_fig_conc_d_3} shows the concentration field generated by a point source and sink dipole. The mobility approximation is less applicable near the source and the sink since the magnitude of $\tilde{c}$ approaches unity. However, the magnitudes of $\tilde{c}$ are significantly smaller away from the source and sink, and our mobility approximation remains valid in most of the region. We note the negative concentration values as the initial concentration was taken to be zero. The values can be offset simply by choosing a different initial condition. The results  will remain unaffected since the particle velocities only rely on the difference of concentrations and are not influenced by the absolute value.
\section*{Conflicts of interest}
There are no conflicts to declare.

\section*{Acknowledgements}
The authors would like to thank Filipe Henrique, Nathan Jarvey, Arkava Ganguly, Dr. Jin Gyun Lee, Cooper Thome, Nicole Day, Taylor Ausec, Kendra Kreienbrink, Gesse Roure, and Dr. Suin Shim for their feedback and insightful discussions leading to the completion of this work. The authors would also like to thank the anonymous referees for their insightful feedback. Ankur Gupta acknowledges support from the American Chemical Society (ACS) Petroleum Research Fund. C. Wyatt Shields IV acknowledges support from the National Science Foundation (NSF) CAREER grant (CBET 2143419).



\balance

\renewcommand\refname{References}

\bibliography{references} 
\bibliographystyle{rsc} 

\end{document}